\documentclass[aps,preprint,showpacs]{revtex4}
\usepackage{amsfonts}
\usepackage{amsmath}
\usepackage{amssymb}
\usepackage{graphicx}

\input{tcilatex}

\begin{document}

\title{Spin-dependent (magneto)transport through a ring due to spin-orbit
interaction}
\author{B. Moln\'{a}r}
\affiliation{Departement Natuurkunde, Universiteit Antwerpen (Campus Drie Eiken), B-2610
Antwerpen, Belgium}
\author{F. M. Peeters}
\affiliation{Departement Natuurkunde, Universiteit Antwerpen (Campus Drie Eiken), B-2610
Antwerpen, Belgium}
\author{P. Vasilopoulos}
\affiliation{Department of Physics, Concordia University, 1455 de Maisonneuve Quest,
Montr\'eal, Quebec, Canada H3G 1M8}

\begin{abstract}
Electron transport through a one-dimensional ring connected with two
external leads, in the presence of spin-orbit interaction (SOI) of strength $%
\alpha$ and a perpendicular magnetic field is studied. Applying Griffith's
boundary conditions we derive analytic expressions for the reflection and
transmission coefficients of the corresponding one-electron scattering
problem. We generalize earlier conductance results by Nitta \textit{et al.}
[Appl. Phys. Lett. \textbf{75}, 695 (1999)] and investigate the influence of 
$\alpha$, temperature, and a weak magnetic field on the conductance. Varying 
$\alpha$ and temperature changes the position of the minima and maxima of
the magnetic-field dependent conductance, and it may even convert a maximum
into a minimum and vice versa.
\end{abstract}

\maketitle

\section{Introduction}

Recently, much attention has been paid to the manipulation of the spin
degrees of freedom of conduction charges in low-dimensional semiconductor
structures. An important feature of the electron transport in multiply
connected systems is that the conductance shows signatures of quantum
interference that depend on the electromagnetic potentials: Aharonov-Bohm
and Aharonov-Casher effect \cite%
{Nitta,AB,AC,Hamilton,Meir,Qian,Oh,Yi,Zhu,Aronov}. A comprehensive review of
results for metallic rings is given in Ref. \cite{web}. Many devices have
been proposed to utilize additional topological phases acquired by the
electrons travelling through quantum circuits \cite%
{Nitta,Euges,Frustaglia,Ionicioiu,Yau}. Nitta \textit{et. al.} proposed a 
\emph{spin-interference} device \cite{Nitta} allowing considerable
modulation of the electric current. This device is a one-dimensional ring
connected with two external leads, made of a semiconductor structure in
which the Rashba spin-orbit interaction (SOI) \cite{Rashba} is the dominant
spin-splitting mechanism. The key idea was that, even in the absence of an
external magnetic field, the difference in the Aharonov-Casher phase \cite%
{AC,Qian} acquired between carriers, travelling clockwise and
counterclockwise, would produce interference effects in the spin-sensitive
electron transport. By tuning the strength $\alpha$ of the SOI the phase
difference could be changed, hence the conductance could be modulated. Nitta 
\textit{et. al.} \cite{Nitta} found that the conductance $G$ is given
approximatively by 
\begin{equation}
G\sim\frac{e^{2}}{h}\ [1+\cos(2\pi\alpha\frac{am^{\ast}}{\hbar^{2}})],
\label{Nittapro}
\end{equation}
where $a$ is the radius of the ring and $m^{\ast}$ the effective mass of the
carriers. It is of interest to verify the validity of this strong sinusoidal
modulation of the conductance, predicted by Eq. (\ref{Nittapro}).

The Rashba field involved in Ref. 1 results from the asymmetric confinement
along the direction ($z$) perpendicular to the plane of the ring. A similar
study but with this field tilted away from
the $z$ direction, by an angle $\phi$, was made in Ref. 17. The
resulting Rashba field is weaker since the radial part of the
confinement is much weaker$^{18}$ but this was not
elaborated in Ref. 17. The  transmission coefficient of Ref. 17
coincides with ours for $\phi=0$ but it is less general in two
important aspects: it is valid only for zero temperature and  in
the absence of a magnetic field whereas ours is free from these
limitations.

In this paper we present an \textit{exact}, \textit{analytic} treatment of
the influence of the SOI on the electron transport through the
spin-interference device of Ref. \cite{Nitta}. Applying Griffith's boundary
conditions \cite{Griffith,Xia} at the junction points we solve the
corresponding scattering problem analytically, obtain the correct form of
the conductance $G$, and show how for large $\alpha$ it is modulated
approximately as predicted by Eq. (\ref{Nittapro}). Further, we assess the
influence of a weak magnetic field on this conductance, indicate the
spin-filtering properties of the ring, and generalize the result to finite
temperatures. 
These latter aspects were not studied at all in Ref. \cite{tae}.

The paper is organized as follows. In Sec. II we solve the one-electron
problem for a ring in the presence of SOI at zero magnetic field and apply
Griffith's boundary conditions. In Sec. III we evaluate in detail the
transmission and reflection coefficients and the zero-temperature
conductance. In Sec. IV we reevaluate the conductance in the presence of a
weak magnetic field and point out the relevance of the results to spin
filtering. In Sec. V we present the finite-temperature conductance and some
numerical results. Concluding remarks follow in Sec. VI 
and details about the spin eigenstates and probability currents are given in
the appendix.\newline

\section{One-electron problem}

\subsection{Hamiltonian}

In the presence of SOI the Hamiltonian operator for a one-dimensional ring
structure is given by \cite{Meijer} 
\begin{equation}
\widehat{H}=-\hbar \Omega \frac{\partial ^{2}}{\partial \varphi ^{2}}-i\hbar
\omega _{so}(\cos \varphi \sigma _{x}+\sin \varphi \sigma _{y})\frac{%
\partial }{\partial \varphi }-i\frac{\hbar \omega _{so}}{2}(\cos \varphi
\sigma _{y}-\sin \varphi \sigma _{x}),  \label{Ha1}
\end{equation}%
where 
$\sigma _{x}$, $\sigma _{y}$ and $\sigma _{z}$ are the Pauli matrices. The
parameter $\Omega $ denotes $\hbar /2m^{\ast }a^{2}$ and $\omega
_{so}=\alpha /\hbar a$ is the frequency associated to the SOI. 
The Rashba field we consider here results from the asymmetric confinement
along the direction ($z$) perpendicular to the plane of the ring. 
The parameter $\alpha $ represents the average electric field along the $z$
direction and is assumed to be a tunable quantity. For an InGaAs-based two
dimensional electron gas, $\alpha $ can be controlled by a gate voltage with
typical values in the range $(0.5-2.0)\times 10^{-11}\mathrm{eVm}$. \cite{22,23} Writing
the Pauli matrices in cylindrical coordinates, 
\begin{equation}
\sigma _{r}=\cos \varphi \sigma _{x}+\sin \varphi \sigma _{y},\hspace{0.55cm}%
\sigma _{\varphi }=\cos \varphi \sigma _{y}-\sin \varphi \sigma _{x},
\end{equation}%
and using $\partial \sigma _{r}/\partial \varphi =\sigma _{\varphi }$ one
can recast the Hamiltonian in the more compact form 
\begin{equation}
\widehat{H}=\hbar \Omega (-i\frac{\partial }{\partial \varphi }+\frac{\omega
_{so}}{2\Omega }\sigma _{r})^{2}\text{.}  \label{Ham2}
\end{equation}%
An  additive constant term $\omega _{so}^{2}/4\Omega $ has been
neglected in Eq. (\ref{Ha1}); as a result it will not appear in the eigenvalues given in Eq. (6a). The neglect is justified for $ka\gg \omega _{so}/2\Omega $. It should be emphasized that this Hamiltonian
is a Hermitian operator \cite{Meijer}, under proper boundary conditions, in
contrast to the non-Hermitian one used in \cite{Nitta}. As can be seen
above, the SOI enters Eq. (\ref{Ham2}) as the spin-dependent vector
potential $(\omega _{so}/2\Omega )\sigma _{r}$. It is convenient to
introduce the dimensionless Hamiltonian 
\begin{equation}
H=\frac{1}{\hbar \Omega }\widehat{H}=(-i\frac{\partial }{\partial \varphi }+%
\frac{\omega _{so}}{2\Omega }\sigma _{r})^{2}.
\end{equation}%
%
%
%
Then, as outlined in the appendix, one can solve the eigenvalue problem in a
straightforward manner. The energy spectrum $E_{n}^{(\mu )}$ and
unnormalized eigenstates $\Psi _{n}^{(\mu )}$ (the normalization depends on
the boundary conditions), labeled by the index $\mu =1,2$, are found to be 
\begin{subequations}
\begin{align}
E_{n}^{(\mu )}& =(n-\Phi _{AC}^{(\mu )}/2\pi )^{2},  \label{spect} \\
\Psi _{n}^{(\mu )}(\varphi )& =e^{in\varphi }\chi _{n}^{(\mu )}(\varphi );
\label{stat}
\end{align}%
here the mutually orthogonal spinors $\chi ^{(\mu )}(\varphi )$ can be
expressed in terms of the eigenvectors $\binom{1}{0},\binom{0}{1}$ of the
Pauli matrix $\sigma _{z}$ as 
\end{subequations}
\begin{subequations}
\begin{align}
\chi _{n}^{(1)}(\varphi )& =\binom{\cos \frac{\theta }{2}}{e^{i\varphi }\sin 
\frac{\theta }{2}},  \label{spin1} \\
\chi _{n}^{(2)}(\varphi )& =\binom{\sin \frac{\theta }{2}}{-e^{i\varphi
}\cos \frac{\theta }{2}},  \label{spin2}
\end{align}%
with the angle $\theta $ given by 
\end{subequations}
\begin{equation}
\theta =2\arctan \left( \Omega -\sqrt{\Omega ^{2}+\omega _{so}^{2}}\right)
/\omega _{so}.  \label{theta}
\end{equation}%
The spin-dependent term $\Phi _{AC}^{(\mu )}$ is the Aharonov-Casher phase

\begin{equation}
\Phi _{AC}^{(\mu )}=-\pi \left[ 1+(-1)^{\mu }\left( \omega _{so}^{2}+\Omega
^{2}\right) ^{1/2}/\Omega \right] .
\end{equation}%
Until now we have not specified the boundary conditions and solved only the
time-independent Schr\"{o}dinger equation. However, it can be seen from Eqs.
(\ref{spect}) and (\ref{stat}) that whatever the boundary conditions, in the
presence of SOI the solution of the Schr\"{o}dinger equation differs from
the unnormalized, free-energy eigenstates only in the phase factor $\exp
(i\varphi \Phi _{AC}^{(\mu )}/2\pi ).$ In words Eq. (\ref{stat}) means that
the unnormalized spinor $\Psi _{n}^{(\mu )}$ picks up the Aharonov-Casher
phase $\Phi _{AC}^{(\mu )}$ upon encircling the ring once.

\subsection{Device geometry and boundary conditions}

The ring connected to two leads is shown in Fig. 1 with the local coordinate
systems attached to the different regions of the device. If the ring is not
connected to any leads the natural boundary condition is that the wave
function has to be single valued when the argument $\varphi$ is increased by
an integral multiple of $2\pi$; this entails that the quantum number $n$
(see Eq. (\ref{stat})) must be integer. Connecting the ring to external
leads alters this condition. In this case it is appropriate to apply a
spin-dependent version of the Griffith's boundary conditions \cite%
{Griffith,Xia} at the intersections as we will specify below. This reduces
the electron transport through the spin-interference device to an exactly
solvable, one-dimensional scattering problem. According to these boundary
conditions at each junction: \textit{i})\textit{\ }the wave function must be
continuous, and \textit{ii}) the spin probability current density must be
conserved.

In the present problem the total wavefunction in the incoming and the
outgoing lead can be expanded in terms of spinors $\chi ^{(\mu )}$ of Eqs. (%
\ref{spin1}) and (\ref{spin2}) as 
\begin{subequations}
\begin{align}
\Psi _{I}(x)& =\underset{\mu =1,2}{\sum }\Psi _{I}^{(\mu )}(x)\chi ^{(\mu
)}(\pi ),\hspace{0.85cm}x\in \left[ -\infty ,0\right] , \\
\Psi _{II}(x^{\prime })& =\underset{\mu =1,2}{\sum }\Psi _{II}^{(\mu
)}(x^{\prime })\chi ^{(\mu )}(0),\hspace{0.85cm}x^{\prime }\in \left[
0,\infty \right] ,
\end{align}%
respectively. (See Fig. 1 for the local coordinates $x$ and $x^{\prime }$.)
The coefficients are the single spin wave functions $\Psi _{I}^{(\mu )}(x)$
and $\Psi _{II}^{(\mu )}(x^{\prime })$ having the from 
\end{subequations}
\begin{subequations}
\begin{align}
\Psi _{I}^{(\mu )}(x)& =(e^{ikx}f_{\mu }+e^{-ikx}r_{\mu }), \\
\Psi _{II}^{(\mu )}(x^{\prime })& =e^{ikx^{\prime }}t_{\mu },
\end{align}%
respectively, where $k$ denotes the incident wave number, $f_{1}=\cos
(\gamma /2)$ and $f_{2}=\sin (\gamma /2)$. As it can be seen, $r_{\mu }$ is
the reflection coefficent while $t_{\mu }$ is the transmisson coefficient
for spin polarization $\mu $ ($\mu =1,2$). In a similar fashion the wave
functions corresponding to the upper and lower arms of the ring can be
written as 
\end{subequations}
\begin{subequations}
\begin{align}
\Psi _{up}(\varphi )& =\underset{\mu =1,2}{\sum }\Psi _{up}^{(\mu )}(\varphi
)\chi ^{(\mu )}(\varphi ),\hspace{0.85cm}\varphi \in \left[ 0,\pi \right] ,
\\
\Psi _{low}(\varphi ^{\prime })& =\underset{\mu =1,2}{\sum }\Psi
_{low}^{(\mu )}(\varphi ^{\prime })\chi ^{(\mu )}(-\varphi ^{\prime }),%
\hspace{0.85cm}\varphi ^{\prime }\in \left[ 0,\pi \right] ,
\end{align}%
respectively (see Fig. 1 for coordinates). The corresponding wave functions
read 
\end{subequations}
\begin{subequations}
\begin{align}
\Psi _{up}^{(\mu )}(\varphi )& =\overset{2}{\underset{j=1}{\sum }}a_{j}^{\mu
}e^{in_{j}^{\mu }\varphi }, \\
\Psi _{low}^{(\mu )}(\varphi ^{\prime })& =\overset{2}{\underset{j=1}{\sum }}%
b_{j}^{\mu }e^{-in_{j}^{\mu }\varphi ^{\prime }}.
\end{align}%
Here the real numbers $n_{j}^{\mu }$ ($j=1,2$) 
\end{subequations}
\begin{equation}
n_{j}^{\mu }=(-1)^{j}ka+\Phi _{AC}^{(\mu )}/2\pi ,
\end{equation}%
are the solutions of the equation $k^{2}a^{2}=E_{n^{\mu }}^{\mu }$ ensuring
the conservation of energy. The coefficients $r_{\mu }$, $t_{\mu }$, $%
a_{j}^{\mu }$ and $b_{j}^{\mu }$ are not independent: they are connected to
each other via Griffith's boundary conditions. First applying the continuity
conditions for the wave functions $\Psi _{II}^{(\mu )}(0)=\Psi _{up}^{(\mu
)}(0)=\Psi _{low}^{(\mu )}(0)$ and $\Psi _{I}^{(\mu )}(0)=\Psi _{up}^{(\mu
)}(\pi )=\Psi _{low}^{(\mu )}(\pi )$, one finds 
\begin{subequations}
\begin{align}
\overset{2}{\underset{j=1}{\sum }}a_{j}^{\mu }& =\overset{2}{\underset{j=1}{%
\sum }}b_{j}^{\mu }=t_{\mu },  \label{E1} \\
\overset{2}{\underset{j=1}{\sum }}a_{j}^{\mu }e^{in_{j}^{\mu }\pi }& =%
\overset{2}{\underset{j=1}{\sum }}b_{j}^{\mu }e^{-in_{j}^{\mu }\pi }=r_{\mu
}+f_{\mu }.  \label{E2}
\end{align}

Now let us turn to the second boundary condition. If one assumes that there
are no spin-flip processes at the junctions, one requires that the spin
probability currents $J^{\mu }$ for each spin direction $\mu $ should be
conserved, \textit{i.e.:} $J_{up}^{\mu }+J_{low}^{\mu }+J_{I(II)}^{\mu }=0$. 
As shown in the appendix, the dimensionless spin currents in the ring arms
are found to be 
\end{subequations}
\begin{subequations}
\begin{align}
J_{up}^{\mu }(\varphi )& =2\func{Re}\{(\Psi _{up}^{(\mu )}\chi ^{(\mu
)})^{\dagger }(-i\partial /\partial \varphi +\omega _{so}\sigma _{r}/2\Omega
)\Psi _{up}^{(\mu )}\chi ^{(\mu )}\}, \\
J_{low}^{\mu }(\varphi ^{\prime })& =\func{Re}\{(\Psi _{low}^{(\mu )}\chi
^{(\mu )})^{\dagger }(-i\partial /\partial \varphi ^{\prime }-\omega
_{so}\sigma _{r}^{\prime }/2\Omega )\Psi _{low}^{(\mu )}\chi ^{(\mu )}\},
\end{align}%
where $\sigma _{r}^{\prime }(\varphi ^{\prime })=\sigma _{r}(\varphi
=-\varphi ^{\prime })=\cos \varphi ^{\prime }\sigma _{x}-\sin \varphi
^{\prime }\sigma _{y}$ because of the orientation of the coordinate system
in the lower arm is opposite to that in the upper arm. The currents in the
leads are given by 
\end{subequations}
\begin{subequations}
\begin{align}
J_{I}^{\mu }(x)& =2a\func{Re}\{(\Psi _{I}^{(\mu )}\chi ^{(\mu )})^{\dagger
}(-i\partial /\partial x)\Psi _{I}^{(\mu )}\chi ^{(\mu )}\}, \\
J_{II}^{\mu }(x^{\prime })& =2a\func{Re}\{(\Psi _{II}^{(\mu )}\chi ^{(\mu
)})^{\dagger }(-i\partial /\partial x^{\prime })\Psi _{II}^{(\mu )}\chi
^{(\mu )}\}.
\end{align}%
Here it should be emphasized that the spinors $\chi ^{(\mu )}$ ($\mu =1,2$)
are obviously the eigenstates of the operator $-i\partial /\partial \varphi
+(\omega _{so}/2\Omega )\sigma _{r}$, which$\ $commutes with $\hat{H}$ given
by Eq. (\ref{Ham2}). Therefore $J^{\mu }$ are well-defined conserved
spin-current densities in the ring. Using the previous requirement $\Psi
_{I(II)}^{(\mu )}=\Psi _{up}^{(\mu )}=\Psi _{low}^{(\mu )}$ at the
junctions, the conservation of the spin current densities can be simply
written as 
\end{subequations}
\begin{equation}
\left. \partial \Psi _{up}^{(\mu )}\right\vert _{\varphi =0(\pi )}+\left.
\partial \Psi _{low}^{(\mu )}\right\vert _{\varphi ^{\prime }=0(\pi
)}+\left. a\partial \Psi _{II(I)}^{(\mu )}\right\vert _{x^{\prime }(x)=0}=0.
\end{equation}%
Evaluating the derivatives, one obtains 
\begin{subequations}
\begin{align}
\overset{2}{\underset{j=1}{\sum }}a_{j}^{\mu }\frac{n_{j}^{\mu }+1/2}{ka}-%
\overset{2}{\underset{j=1}{\sum }}b_{j}^{\mu }\frac{n_{j}^{\mu }+1/2}{ka}%
+t_{\mu }& =0,  \label{G1} \\
\overset{2}{\underset{j=1}{\sum }}a_{j}^{\mu }e^{in_{j}^{\mu }\pi }\frac{%
n_{j}^{\mu }+1/2}{ka}-\overset{2}{\underset{j=1}{\sum }}b_{j}^{\mu
}e^{-in_{j}^{\mu }\pi }\frac{n_{j}^{\mu }+1/2}{ka}+f_{\mu }-r_{\mu }& =0.
\label{G2}
\end{align}%
The variables $r_{\mu }$, $t_{\mu }$ can be eliminated using Eqs. (\ref{E1})
and (\ref{E2}). Then the set of Eqs. (\ref{G1}) and (\ref{G2}) is replaced
by the linear set of algebraic equations for the coefficients $\left\{
a_{j}^{\mu },b_{j}^{\mu }\right\} $: 
\end{subequations}
\begin{subequations}
\begin{align}
\overset{2}{\underset{j=1}{\sum }}a_{j}^{\mu }\frac{n_{j}^{\mu }+ka}{ka}-%
\overset{2}{\underset{j=1}{\sum }}b_{j}^{\mu }\frac{n_{j}^{\mu }}{ka}& =0,
\label{G3} \\
\overset{2}{\underset{j=1}{\sum }}a_{j}^{\mu }e^{in_{j}^{\mu }\pi }\frac{%
n_{j}^{\mu }-ka}{ka}-\overset{2}{\underset{j=1}{\sum }}b_{j}^{\mu
}e^{-in_{j}^{\mu }\pi }\frac{n_{j}^{\mu }}{ka}& =-2f_{\mu }.  \label{G4}
\end{align}

\section{Transmission and reflection coefficients, conductance}

The linear equations (\ref{E1}) and (\ref{E2}) together with (\ref{G3}) and (%
\ref{G4}) for the variables $a_{j}^{\mu }$ and $b_{j}^{\mu }$ can be written
in the matrix form: 
\end{subequations}
\begin{equation}
M^{\mu }\left[ 
\begin{array}{c}
a_{1}^{\mu } \\ 
a_{2}^{\mu } \\ 
b_{1}^{\mu } \\ 
b_{2}^{\mu }%
\end{array}%
\right] =-2\left[ 
\begin{array}{c}
0 \\ 
0 \\ 
0 \\ 
f_{\mu }%
\end{array}%
\right] ,
\end{equation}%
with the matrix $M^{\mu }$ depending only on the wave number $ka$ and $%
n_{j}^{\mu }$: 
\begin{equation}
M^{\mu }=\left[ 
\begin{array}{cccc}
1 & 1 & -1 & -1 \\ 
e^{in_{1}^{\mu }\pi } & e^{in_{2}^{\mu }\pi } & -e^{-in_{1}^{\mu }\pi } & 
-e^{-in_{2}^{\mu }\pi } \\ 
\frac{n_{1}^{\mu }+ka}{ka} & \frac{n_{2}^{\mu }+ka}{ka} & -\frac{n_{1}^{\mu }%
}{ka} & -\frac{n_{2}^{\mu }}{ka} \\ 
\frac{n_{1}^{\mu }-ka}{ka}e^{in_{1}^{\mu }\pi } & \frac{n_{2}^{\mu }-ka}{ka}%
e^{in_{2}^{\mu }\pi } & -\frac{n_{1}^{\mu }}{ka}e^{-in_{1}^{\mu }\pi } & -%
\frac{n_{2}^{\mu }}{ka}e^{-in_{2}^{\mu }\pi }%
\end{array}%
\right] \text{.}
\end{equation}%
Now let us calculate the transmission ($t_{\mu }$) and reflection ($r_{\mu }$%
) coefficients which are connected to the incoming spinor according to the
following equations: 
\begin{subequations}
\begin{align}
\left( 
\begin{array}{c}
t_{1} \\ 
t_{2}%
\end{array}%
\right) & =T\left( 
\begin{array}{c}
\cos \frac{\gamma }{2} \\ 
\sin \frac{\gamma }{2}%
\end{array}%
\right) =\left[ 
\begin{array}{cc}
T_{1} & 0 \\ 
0 & T_{2}%
\end{array}%
\right] \left( 
\begin{array}{c}
\cos \frac{\gamma }{2} \\ 
\sin \frac{\gamma }{2}%
\end{array}%
\right) , \\
\left( 
\begin{array}{c}
r_{1} \\ 
r_{2}%
\end{array}%
\right) & =R\left( 
\begin{array}{c}
\cos \frac{\gamma }{2} \\ 
\sin \frac{\gamma }{2}%
\end{array}%
\right) =\left[ 
\begin{array}{cc}
R_{1} & 0 \\ 
0 & R_{2}%
\end{array}%
\right] \left( 
\begin{array}{c}
\cos \frac{\gamma }{2} \\ 
\sin \frac{\gamma }{2}%
\end{array}%
\right) .
\end{align}%
Both diagonal matrices $T$ and $R$ can be expressed in terms of the inverse
of the $4\times 4$ matrix $M^{\mu }$ in the manner 
\end{subequations}
\begin{subequations}
\begin{align}
T_{\mu }& =-2\left[ (M^{\mu })_{1,4}^{-1}+(M^{\mu })_{2,4}^{-1}\right] ,%
\hspace{0.1cm}  \label{trans} \\
R_{\mu }& =-2\left[ e^{in_{1}^{\mu }\pi }(M^{\mu
})_{1,4}^{-1}+e^{in_{2}^{\mu }\pi }(M^{\mu })_{2,4}^{-1}+1/2\right] .
\end{align}%
Calculating the $4^{\mathrm{th}}$ row of the inverse matrix gives 
\end{subequations}
\begin{subequations}
\begin{align}
T_{\mu }& =-8i\cos (\Theta _{\mu }\pi )\sin (\Lambda _{\mu }\pi )/d_{\mu },
\label{ftrans} \\
R_{\mu }& =[\cos (2\Lambda _{\mu }\pi )-1]ka/\Lambda _{\mu }d+4[\cos
(2\Theta _{\mu }\pi )-\cos (2\Lambda _{\mu }\pi )]\Lambda _{\mu }/kad_{\mu },
\end{align}%
with the following notations 
\end{subequations}
\begin{subequations}
\begin{gather}
d_{\mu }=\left[ \cos (2\Lambda _{\mu }\pi )-1\right] ka/\Lambda _{\mu }+4%
\left[ \cos (2\Lambda _{\mu }\pi )-\cos (2\Theta _{\mu }\pi )\right] \Lambda
_{\mu }/ka-4i\sin (2\Lambda _{\mu }\pi )\text{,} \\
\Lambda _{\mu }=\left( n_{2}^{\mu }-n_{1}^{\mu }\right) /2,\text{\hspace{%
1.05cm}}\Theta _{\mu }=\left( n_{2}^{\mu }+n_{1}^{\mu }\right) /2.
\end{gather}%
One can verify that for each spin polarization $\mu $ ($\mu =1,2$): 
\end{subequations}
\begin{equation}
\left\vert T_{\mu }\right\vert ^{2}+\left\vert R_{\mu }\right\vert ^{2}=1.
\end{equation}%
Here we would like to point out that the expressions for $T_{\mu }$ and $%
R_{\mu }$ above are quite general. They are still valid for other
Hamiltonians than the one used, provided the spinors $\chi _{n_{1}^{\mu }}$
and $\chi _{n_{2}^{\mu }}$, which travel clockwise and counterclockwise,
respectively, are along the same direction.

In the present case $\Lambda _{\mu }=ka$ and $\Theta _{\mu }=\Phi
_{AC}^{(\mu )}/2\pi $. Consequently the concrete expression for the
transmission amplitudes reads 
\begin{equation}
T_{\mu }=\frac{8i\cos (\Phi _{AC}^{(\mu )}/2)\sin (ka\pi )}{1-5\cos (2ka\pi
)+4\cos \Phi _{AC}^{(\mu )}+4i\sin (2ka\pi )}.  \label{tlast}
\end{equation}%
In the Landauer formalism the conductance is given by 
\begin{equation}
G=\frac{e^{2}}{h}\underset{\mu ,\lambda =1}{\overset{2}{\sum }}\left\vert
T_{\mu \lambda }\right\vert ^{2}.  \label{Lanform}
\end{equation}%
In the present case the off-diagonal elements $T_{12}$ and $T_{21}$ of the
transmission matrix are zero. Inserting Eq. (\ref{tlast}) in Eq. (\ref%
{Lanform}) we obtain the \textit{exact} conductance at zero temperature in
the form 
\begin{equation}
G=(e^{2}/h)g_{0}(k,\Delta _{AC})[1-\cos (\Delta _{AC})],  \label{totco1}
\end{equation}%
where the dimensionless coefficent $g_{0}$ is 
\begin{equation}
g_{0}(k,\Delta _{AC})=\frac{64\sin ^{2}(ka\pi )}{[1-5\cos (2ka\pi )-4\cos
(\Delta _{AC})]^{2}+16\sin ^{2}(2ka\pi )}.  \label{Con1}
\end{equation}%
Here $\Delta _{AC}=(\Phi _{AC}^{(1)}-\Phi _{AC}^{(2)})/2=\pi \lbrack
(2m^{\ast }a/\hbar ^{2})^{2}\alpha ^{2}+1]^{1/2}$ is the half of the
difference between the phases accumulated by the different spinors.
Comparing Eq. (\ref{totco1}) with the approximate formula (1) one can see
that the conductance oscillates with $\cos (\Delta _{AC})$ in a more complex
manner. For large values of the Rashba parameter $\alpha $ an essential
difference is a $\pi $ phase shift in the oscillation; however, the period
remains the same. An important feature is the presence of the factor $\cos
(\Delta _{AC})$ in the denominator of Eq. (\ref{Con1}). This makes $g_{0}$
not a constant equal to $1$, as found in Ref. \cite{Nitta}, but a quantity
that depends on $\Delta _{AC}$ and the incident energy through $k$. The full
dependence of $g_{0}$ on $\Delta _{AC}$ for different temperatures,
including $T=0$, is shown in Sec. V.

Figure 2 shows the conductance $G$ versus $\Delta _{AC}$ at different wave
numbers $k$. Because $G$ is an even and periodic function of $ka$ (with
period $1$), it is sufficient to consider only the half period $ka\in
\lbrack 0,1/2]$. One can see that $k\approx 0$ (or for $ka\approx l\in 
\mathbb{N}$) the conductance tends to a discontinuous function which is
non-zero only at $\Delta _{AC}=\pi +2n\pi $ ($n$ is integer) with value $%
2e^{2}/h$. This dependence of $G$ on $ka$ is absent in Eq. (\ref{Nittapro}). 
We note in passing that a transmission coefficent formally equivalent to Eq.
(\ref{totco1}) was derived earlier in Ref. \cite{tae} with very few details
and starting with a Hamiltonian in which  the Rashba field is tilted away 
from the $z$ direction by an angle $\phi$. It
coincides with ours  for $\phi=0$. As shown below, however, ours
takes into account finite temperatures and a weak magnetic field
whereas that of Ref. 17 does not. In addition, we give the reflection
coefficient in Eq. (25b).

\section{Weak magnetic perturbation}

Our analytic result can be easily extended to the case of a weak magnetic
perturbation. Let us suppose that an external magnetic field $%
\overrightarrow{B}$ normal to the plane of the ring is present. Then the
vector potential can be chosen to be tangential 
\begin{equation}
\overrightarrow{A}=(Ba/2)\overrightarrow{e}_{\varphi }.
\end{equation}%
First we take the effect of the magnetic flux $\Phi =a\oint \overrightarrow{A%
}d\overrightarrow{\varphi }$ encircled by the ring into consideration. It
means that we have to change the momentum operator $-i\hbar \nabla $ in the
Hamiltonian with $-i\hbar \nabla -e\overrightarrow{A}$ ('minimal coupling'
substitution). This leads to the appearance of the magnetic flux $\Phi /\Phi
_{0}$ in the Hamiltonian, where $\Phi _{0}=h/e$ is the unit of flux, if the
Zeeman term $g^{\ast }\overrightarrow{B}\cdot \overrightarrow{S}$ is
neglected \cite{Meir}. Then the Aharonov-Bohm phase picked up by an electron
encircling this magnetic flux 
\begin{equation}
\Phi _{AB}=2\pi \Phi /\Phi _{0}=\pi eBa^{2}/\hbar ,
\end{equation}%
and the dimensionless Hamiltonian in question reads: 
\begin{equation}
H=(-i\frac{\partial }{\partial \varphi }-\frac{\Phi _{AB}}{2\pi }-\frac{%
\omega _{so}}{2\Omega }\sigma _{r})^{2}.  \label{Hammag}
\end{equation}%
When the Zeeman term is present, the interaction between the electron spin
and a relatively \textit{weak} magnetic field $B$ can be treated by
perturbation theory. Using the dimensionless field strength $b=g^{\ast
}eB/4m\Omega $ the perturbation of the Hamiltonian (\ref{Hammag}) is 
\begin{equation}
H_{p}=b\sigma _{z}=(g^{\ast }m^{\ast }/m)\Phi _{AB}\sigma _{z},
\end{equation}%
where $m$ is the bare electron mass and $g^{\ast }$ the effective
gyromagnetic ratio. The matrix elements of $H_{p}$ in the basis of the
normalized eigenstates $\left\vert \Psi _{n}^{\mu }\right\rangle =\Psi
_{n}^{\mu }(\varphi )/\sqrt{2\pi }$ are obtained as 
\begin{subequations}
\begin{align}
\left\langle \Psi _{n}^{\mu }\right\vert H_{p}\left\vert \Psi _{n}^{\mu
}\right\rangle & =(-1)^{\mu +1}(g^{\ast }m^{\ast }/m)\Phi _{AB}\cos
\vartheta =(-1)^{\mu +1}C_{\vartheta }, \\
\left\langle \Psi _{n}^{1}\right\vert H_{p}\left\vert \Psi
_{n}^{2}\right\rangle & =(g^{\ast }m^{\ast }/m)\Phi _{AB}\sin \vartheta .
\end{align}%
In the first-order approximation one neglects the off-diagonal elements;
this is reasonable if they are small, \textit{i.e.}, if $(g^{\ast }m^{\ast
}/m)\Phi _{AB}\ll k^{2}a^{2}$. To first order the eigenspinors are not
perturbed and their direction is still specified by the angle $\vartheta $
given by Eq. (\ref{theta}). Using the identity 
\end{subequations}
\begin{equation}
\cos \vartheta =\frac{1-\tan ^{2}(\vartheta /2)}{1+\tan ^{2}(\vartheta /2)}%
=\pi /\Delta _{AC},
\end{equation}%
we obtain the energies, including the first-order corrections, 
\begin{equation}
E_{n}^{\mu }=(n-\frac{\Phi _{AB}}{2\pi }-\frac{\Phi _{AC}^{(\mu )}}{2\pi }%
)^{2}-(-1)^{\mu }\frac{g^{\ast }m^{\ast }\pi }{m\Delta _{AC}}\Phi _{AB}.
\end{equation}%
The equation of energy conservation $k^{2}a^{2}=E_{n^{\mu }}^{\mu }$ has the
solutions 
\begin{subequations}
\begin{align}
n_{1}^{\mu }& =-\sqrt{k^{2}a^{2}+(-1)^{\mu }C_{\vartheta }}+\Phi _{AB}/2\pi
+\Phi _{AC}^{(\mu )}/2\pi , \\
n_{2}^{\mu }& =\sqrt{k^{2}a^{2}+(-1)^{\mu }C_{\vartheta }}+\Phi _{AB}/2\pi
+\Phi _{AC}^{(\mu )}/2\pi .
\end{align}%
Because the eigenspinors are not modified within this approximation the
transmission matrix elements are given again by Eq. (\ref{ftrans}) but with
the parameters $\Lambda _{\mu }$ and $\Theta _{\mu }$ replaced, repectively
by 
\end{subequations}
\begin{subequations}
\begin{equation*}
\Lambda _{\mu }=(n_{2}^{\mu }-n_{1}^{\mu })/2=\sqrt{k^{2}a^{2}+(-1)^{\mu
}C_{\vartheta }},
\end{equation*}%
and 
\begin{equation*}
\Theta _{\mu }=(n_{2}^{\mu }+n_{1}^{\mu })/2=\Phi _{AB}/2\pi +\Phi
_{AC}^{(\mu )}/2\pi =\Phi /\Phi _{0}+\Phi _{AC}^{(\mu )}/2\pi .
\end{equation*}%
This leads to the transmission coefficient 
\end{subequations}
\begin{equation}
T_{\mu }=\frac{8i\cos \left( \Phi _{AB}/2+\Phi _{AC}^{(\mu )}/2\right) \sin
(C_{\mu }ka\pi )}{C_{\mu }^{-1}-(C_{\mu }^{-1}+4)\cos (2C_{\mu }ka\pi
)+4\cos \left( \Phi _{AB}+\Phi _{AC}^{(\mu )}\right) +4i\sin (2C_{\mu }ka\pi
)},  \label{trans2}
\end{equation}%
where $C_{\mu }=\sqrt{1+(-1)^{\mu }C_{\vartheta }/k^{2}a^{2}}$. The
resulting magnetoconductance reads 
\begin{equation}
G=\tfrac{e^{2}}{h}\left[ \left\vert T_{1}\right\vert ^{2}+\left\vert
T_{2}\right\vert ^{2}\right] .  \label{1cond}
\end{equation}

At this point one can envisage an application of the device as a spin
filter. Assuming one can tune the phases $\Phi_{AB}$ and $\Phi_{AC}^{(\mu)}$
(via the magnetic field and the Rashba strength $\alpha$) independently, one
can make the ring almost transparent with high transmission probability only
for electrons with spin quantum number $\mu=1$ ($2$) and totally opaque with 
$\mu=2$ ($1$). For instance, if one sets $\Phi_{AB}+\Phi_{AC}^{(1)}$ and $%
\Phi_{AB}+\Phi_{AC}^{(2)}$ to be $(2p+1)\pi$ and $2q\pi$ into (\ref{trans2}%
), where $q$ and $p$ are integers, respectively, one obtains 
\begin{align}
\left| T_{1}\right| ^{2} & =0, \\
\left| T_{2}\right| ^{2} & =\left( 1+\frac{1+8C}{32C^{2}}\ [1-\cos
(2C_{2}ka\pi)]\right) ^{-1}.  \notag
\end{align}
As can be seen, $\left| T_{2}\right| ^{2}$ has maxima equal to $1$ and
minima equal to $(1+1/4C_{2})^{-2}$ at integer and half-integer values of $%
C_{2}ka$, respectively. Due to the inequality $(g^{\ast}m^{\ast}/m)\Phi
_{AB}\ll k^{2}a^{2}$ we have $C_{2}\sim1$; hence the efficiency of the
filtering process is higher than $64\%$.

\section{Temperature dependence of the conductance}

\subsection{Explicit expression}

The conductance at finite temperatures is given by 
\begin{equation}
G(T)=-\frac{e^{2}}{h}\underset{\mu=1,2}{\sum}\underset{0}{\overset{\infty }{%
\int}}dE\frac{\partial f(E,\mu,T)}{\partial E}\left| T_{\mu}(E)\right| ^{2},
\end{equation}
where $f(E,\mu,T)$ is the Fermi function, $T$ is the temperature and $%
T_{\mu}(E)$ is the single spin-transmission coefficient. In the absence of
magnetic field the conductance can be written as: 
\begin{equation}
G=(e^{2}/h)g_{T}(k_{F},\Delta_{AC})(1-\cos\Delta_{AC}),
\end{equation}
where the explicit form of the temperature depending coefficient $%
g_{T}(k_{F},\Delta_{AC})$ is given by 
\begin{equation}
g_{T}(k_{F},\Delta_{AC})=\underset{0}{\overset{\infty}{\int}}d\zeta \zeta\ 
\frac{(32T_{F}/T)\cosh^{-2}\left[ \left( \zeta^{2}-\overline{\mu }\right)
T_{F}/2T\right] \sin^{2}(\zeta k_{F}a\pi)}{\left[ 1-5\cos(2\zeta
k_{F}a\pi)-4\cos\left( \Delta_{AC}\right) \right] ^{2}+16\sin^{2}(2\zeta
k_{F}a\pi)}.  \label{expform}
\end{equation}
Here $\overline{\mu}$ is the (dimensionless) chemical potential in units of
the Fermi energy $E_{F}$ and $T_{F}$ denotes the Fermi temperature. At $T=0$
the derivative of the Fermi function becomes a $\delta$-function, the
integration in Eq. (\ref{expform}) can be carried out, and one obtains the
previous result $g_{0}$ (Eq. (\ref{Con1})).

In the present of a weak magnetic field ($C_{\mu}\approx1$) the
magnetoconductance reads 
\begin{equation}
G=\frac{e^{2}}{2h}\underset{\mu=1,2}{\sum}g_{T}(k_{F},\Phi_{AB}+(-1)^{\mu
}\Delta_{AC})\left[ 1-\cos(\Phi_{AB}+(-1)^{\mu}\Delta_{AC})\right].
\label{lastmcon}
\end{equation}
As can be seen, the total magnetoconductance for weak fields is the sum of
the two single spin magnetoconductances having the same functional form $%
(e^{2}/2h)g_{T}(k_{F},\phi)\left[ 1-\cos\phi\right] $ but due to the
presence of the SOI they are shifted by the spin-depending phase $%
\pm\Delta_{AC}$ according to Eq. (\ref{lastmcon}).

\subsection{Numerical results}

To stress the difference between our result and the one of Ref. 1 we plot in
Fig. 3(a) and Fig. 3(b) the coefficient $g_{0}$ for different values of $ka$
as indicated. As shown, the coefficient $g_{0}(\Delta_{AC})$ varies in a
rather large range, $0$ up to $16$, depending on the value of $ka$. The
largest deviations from $1$ occur at the end of the period $%
\Delta_{AC}/\pi=2 $ and $4$. Agreement with Ref. 1 is obtained only for
values $\Delta_{AC}/\pi $ in the neighborhood of $3$. This range is the
widest (approximately between $2.5$ and $3.5$) for $ka$ half-integer. For $%
ka $ integer this range collapses into one single point because with this
wave number the coefficient $g_{0}$ is discontinuous having the value $1 $
only at $\Delta_{AC}/\pi=3$ 
and otherwise zero.

In Figs. 4(a) -- 4(d) we investigate the temperature dependence of the
amplitude $g_{T}$ of the oscillations for different values of wave number $%
k_{F}a=20$, $20.25$ and $20.5$. The temperature is expressed in units of the
Fermi temperature $T_{F}$. As seen, for $k_{F}a$ half-integer raising the
temperature reduces the value of $g_{T}$; however for values $k_{F}a$ closer
to an integer the coefficient $g_{T}$ increases until its peaks reach 
a value around $4$. This happens for temperature $T\approx 0.05T_{F}$; and
as one can see, by then the dependence on the fractional part of $k_{F}a$
has already been washed out, too. For a ring of radius $a=0.25\mathrm{\mu m}$
and a Fermi wave number $k_{F}=20.5/a$, with the effective mass of InAs: $%
m^{\ast}=0.023$, 
the Fermi energy $E_{F}$ and the Fermi temperature $T_F$ are $11.13\mathrm{%
meV}$ and $129.27\mathrm{K}$, respectively. With this choice of parameters $%
T=0.05T_{F}$ above corresponds to $6.46\mathrm{K}$. Further increasing the
temperature, now by larger steps, we find that $g_{T}$ decreases much more
slowly.

For the sake of completeness, in Figs. 5(a), 5(b) and 5(c) we present the 
conductance $G=(e^{2}/h)g_{T}\left[ 1-\cos(\Delta_{AC})\right] $ for the
same temperatures and 
values of $k_{F}a$ as in Fig. 4. One can see that by increasing the
temperature the ''camel hump'' like pattern for $k_{F}a$ around
half-integers disappears and $G$ becomes less sensitive to the fractional
part of $k_{F}a$. A more complete dependence of the conductance on $ka$ and $%
\alpha$ at zero temperature is shown in Fig. 6. As can be inferred, \textit{%
e.g.}, by moving along lines of constant $\alpha$ or $ka$, the conductance
depends in a complex manner on $\alpha$ and $ka$. Note that the dependence
of the conductance on $ka$ is completely absent in Eq. (\ref{Nittapro}).

Figures 7(a) and 7(b) show the oscillations of the magnetoconductance versus
magnetic field $B$ in units of $B_{0}=\Phi _{0}/(a^{2}\pi )=21.06\mathrm{mT}$%
, for various values of the SOI strength $\alpha $ and for fixed Fermi wave
number $k_{F}a=20.5$ at $T=0.001T_{F}$ and $T=0.05T_{F}$, respectively. In
both figures the values of $\alpha $ were chosen such that with the above
parameters $m^{\ast }$ and $a$ they correspond to an Aharonov-Casher phase
shift $\Delta _{AC}$ equal to $5\pi /4$, $3\pi /2\ $and $2\pi $ for $\alpha
=0.497\alpha _{0}$, $0.741\alpha _{0}$, and $1.148\alpha _{0}$,
respectively, with $\alpha _{0}=10^{-11}\mathrm{eVm}$. One can see that the
presence of SOI can alter the period of the oscillations, which in its
absence is equal to $B_{0}$ \cite{web}.

In order to get better insight into the positions of extrema in the
magnetoconductance we plotted in Figs. 8(a) and 8(b) those positions as
function of the SOI strength $\alpha $ for fixed temperatures (a) $%
T=0.001T_{F}$ and (b) $T=0.05T_{F}$, respectively. Comparing the figure at
large temperature with the one at low teperature, it can be seen that the
additional substructure of two maxima and a minimum, which is present at $%
T=0.001T_{F}$ and connected with \textquotedblright the camel
hump\textquotedblright\ pattern of the magnetoconductance oscillation, has
been contracted into a single maximum. Further, at both temperatures, near
certain values of $\alpha $, minima (maxima) disappear, and instead of them,
a new maximum (minimum) appears, in other words a bifurcation occurs, in the
oscillation of the magnetoconductance at $B=0,B_{0}/2,$ and $B_{0}$. These
intersections of maximum and minimum curves correspond to saddle points on
the surface of the conductance $G$ depending on both $B$ and $\alpha $. To
show more clearly how changing the strength $\alpha $ can convert a minimum
(maximum) to a maximum (minimum), we plot in Figs. 9(a) and 9(b) the
magnetoconductance in the neighborhood of two saddle points for temperatures 
$T=0.001T_{F}$ and $T=0.05T_{F}$, respectively. For instance, in Fig. 9(b)
one can see that 
for a relatively small increase (decrease) in $\alpha $ around $0.40\alpha
_{0}$ ($1.02\alpha _{0}$) a minimum turns into a maximum surrounded with two
minima at $B=B_{0}$ and $B=B_{0}/2$.

\section{Concluding remarks}

We derived an exact expression for the zero-temperature conductance of a
one-dimensional ring connected to two leads in the presence of SOI. In
addition, we generalized the result to finite temperatures and \textit{weak}
magnetic fields for which the Zeeman term can be treated by perturbation
theory. Since we used the Landauer-B\"uttiker formalism, the conductance
expressions are valid in the ballistic regime.

As specified in the text, the zero-temperature conductance is not as simple
as presented in Ref. 1. Apart from the phase shift $\pi$ between the two
expressions, \textit{cf.} Eqs. (\ref{Nittapro}) and (\ref{totco1}), the
quantity $g_{0}$ is not equal to $1$, as deduced from Eq. (\ref{Nittapro}),
but depends on the strength $\alpha$ of the SOI, on the incident energy, and
the temperature, \textit{cf.} Sec. V. We attribute this difference to the
non-hermitian Hamiltonian and also to the boundary conditions used in Ref.
1. However, the sinusoidal dependence of $G$ on $\alpha$ as predicted in
Ref. 1 is recovered by our exact expression only in the limit of large
values of $\alpha$.

The results presented here are valid for a strictly one-dimensional ring.
They can be extended to rings of finite width $w$ provided the inequality $%
w\ll a$ holds and an infinite-wall confinement is assumed along the radial
direction. In this case the radial and angular motion are decoupled and the
energy levels are shifted by $\hbar^{2}l^{2}/2m^{\ast}w^{2}$, where $l$ is
an integer. The results presented in our paper correspond then to the lowest 
$l=1$ mode.

\begin{acknowledgments}
This work was supported by the Belgian Interuniversity Attraction Poles
(IUAP), the Flemish Concerted Action (GOA) Programme, the Flemish Science
Foundation (FWO-Vl), the EU-CERION programme, the Flemish-Hungarian
Bilateral Programme and by the Canadian NSERC Grant No. OGP0121756. One of
us (B. M.) is supported by a DWTC fellowship to promote S \& T collaboration
between Central and Eastern Europe.\newpage
\end{acknowledgments}

\appendix{\textbf{Appendix} }

Below we give some details of the derivation of the unnormalized eigenstates 
$\Psi_n^{\mu}$, Eq. (6b), and of the spin probability currents in Eqs. (16a)
and (16b).

i) Eigenfunctions $\Psi _{n}^{(\mu )}(\varphi )$.

It is sufficient to solve the eigenvalue problem $H\Psi (\varphi )=E\Psi
(\varphi )$,%
\begin{equation*}
(-i\partial /\partial \varphi +\omega _{so}\sigma _{r}/2\Omega )\Psi
(\varphi )=\Lambda \Psi (\varphi ),
\end{equation*}%
with nergy eigenvalue $E=\Lambda ^{2}$. Writing $\Psi (\varphi )$ in the form%
\begin{equation*}
\Psi (\varphi )=e^{in\varphi }\chi (\varphi )=e^{in\varphi }\left( 
\begin{array}{c}
a \\ 
be^{i\varphi }%
\end{array}%
\right) ,
\end{equation*}%
we obtain%
\begin{equation*}
(-i\partial /\partial \varphi +\omega _{so}\sigma _{r}/2\Omega )\chi
(\varphi )=(\Lambda -n)\chi (\varphi ).
\end{equation*}%
Using $\sigma _{r}=\left( 
\begin{array}{cc}
0 & e^{-i\varphi } \\ 
e^{+i\varphi } & 1%
\end{array}%
\right) $ (in the basis $\binom{1}{0},\binom{0}{1}$ of the eigenstates of
the Pauli matrix $\sigma _{z}$) we obtain%
\begin{equation*}
\left( 
\begin{array}{cc}
0 & \omega _{so}/2\Omega  \\ 
\omega _{so}/2\Omega  & 1%
\end{array}%
\right) \left( 
\begin{array}{c}
a \\ 
b%
\end{array}%
\right) =(\Lambda -n)\left( 
\begin{array}{c}
a \\ 
b%
\end{array}%
\right) .
\end{equation*}%
The eigenvalues of the latter equation are $1/2+(-1)^{\mu }\sqrt{1/4+\omega
_{so}^{2}/4\Omega ^{2}}=-\Phi _{AC}^{(\mu )}/2\pi $, where $\mu =1,2$. The
coefficients of the corresponding eigenvectors can be chosen as $a_{1}=\cos
\theta $, $b_{1}=\sin \theta $,\thinspace $a_{2}=\sin \theta $ and $%
b_{2}=-\cos \theta $, with $\tan \theta =[1/2-\sqrt{1/4+\omega
_{so}^{2}/4\Omega ^{2}}](2\Omega /\omega _{so})=[\Omega -\sqrt{\Omega
^{2}+\omega _{so}^{2}}]/\omega _{so}$. The resulting energy eigenvalues and
unnormalized eigenfunctions are given, respectively, by Eqs. (6a) and (6b).

ii) Spin probability currents

a) We denote a two-component spinor by $\Psi =\left( 
\begin{array}{c}
\Psi _{1} \\ 
\Psi _{2}%
\end{array}%
\right) $ and its complex conjugate by $\overline{\Psi }$. Further, we
introduce the bilinear product by $(\Phi ,\Psi )=\Phi _{1}\Psi _{1}+\Phi
_{2}\Psi _{2}$. Notice that \textit{this is not a scalar product of the
Hilbert space}. One can show that the following continuity equation is valid
for the spinor $\Psi $ obeying the Schr\"{o}dinger equation $i\partial \Psi
/\partial t=H\Psi $ with $H$ given by Eq. (5),

\begin{equation*}
\frac{\partial \rho }{\partial t}+\frac{\partial J}{\partial \varphi }=0;
\end{equation*}%
the probability density is $\rho =(\overline{\Psi },\Psi )$ and the
probability current density $J=2\func{Re}\{(\overline{\Psi },(-i\partial
\Psi /\partial \varphi +(\omega _{so}/2\Omega )\sigma _{r}\Psi )\}$.

\textit{Proof:}

We start with the Schr\"{o}dinger equation $i\partial \Psi /\partial t=H\Psi 
$ written explicitly as%
\begin{equation*}
i\frac{\partial \Psi }{\partial t}=-\frac{\partial ^{2}\Psi }{\partial
^{2}\varphi }-i\frac{\omega _{so}}{\Omega }\sigma _{r}\frac{\partial \Psi }{%
\partial \varphi }-i\frac{\omega _{so}}{2\Omega }\frac{\partial \sigma _{r}}{%
\partial \varphi }\Psi +\frac{\omega _{so}^{2}}{4\Omega ^{2}}\Psi ,
\end{equation*}%
take its and complex conjugate, and consider the products $(\overline{\Psi }%
,H\Psi )$ and $(\overline{H\Psi },\Psi )$%
\begin{equation*}
(\overline{\Psi },H\Psi )=-(\overline{\Psi },\frac{\partial ^{2}\Psi }{%
\partial ^{2}\varphi })-i\frac{\omega _{so}}{\Omega }(\overline{\Psi }%
,\sigma _{r}\frac{\partial \Psi }{\partial \varphi })-i\frac{\omega _{so}}{%
2\Omega }(\overline{\Psi },\frac{\partial \sigma _{r}}{\partial \varphi }%
\Psi )+\frac{\omega _{so}^{2}}{4\Omega ^{2}}(\overline{\Psi },\Psi ),
\end{equation*}%
\begin{equation*}
(\overline{H\Psi },\Psi )=-(\frac{\partial ^{2}\overline{\Psi }}{\partial
^{2}\varphi },\Psi )+i\frac{\omega _{so}}{\Omega }(\overline{\sigma }_{r}%
\frac{\partial \overline{\Psi }}{\partial \varphi },\Psi )+i\frac{\omega
_{so}}{2\Omega }(\frac{\partial \overline{\sigma _{r}}}{\partial \varphi }%
\overline{\Psi },\Psi )+\frac{\omega _{so}^{2}}{4\Omega ^{2}}(\overline{\Psi 
},\Psi ).
\end{equation*}%
Using the fact $(\overline{\sigma _{r}\Phi },\Psi )=(\overline{\Phi },\sigma
_{r}\Psi )$ the latter product can be written as%
\begin{equation*}
(\overline{H\Psi },\Psi )=-(\frac{\partial ^{2}\overline{\Psi }}{\partial
^{2}\varphi },\Psi )+i\frac{\omega _{so}}{\Omega }(\frac{\partial \overline{%
\Psi }}{\partial \varphi },\sigma _{r}\Psi )+i\frac{\omega _{so}}{2\Omega }(%
\overline{\Psi },\frac{\partial \sigma _{r}}{\partial \varphi }\Psi )+\frac{%
\omega _{so}^{2}}{4\Omega ^{2}}(\overline{\Psi },\Psi ).
\end{equation*}%
The derivative $\partial \rho /\partial t$ is given by $(\partial \overline{%
\Psi }/\partial t,\Psi )+(\overline{\Psi },\partial \Psi /\partial t)=i\{(%
\overline{H\Psi },\Psi )-(\overline{\Psi },H\Psi )\}$. Therefore $\partial
\rho /\partial t$ can be written as%
\begin{equation*}
\frac{\partial \rho }{\partial t}=i\{(\overline{\Psi },\frac{\partial
^{2}\Psi }{\partial ^{2}\varphi })-(\frac{\partial ^{2}\overline{\Psi }}{%
\partial ^{2}\varphi },\Psi )+i\frac{\omega _{so}}{\Omega }(\overline{\Psi }%
,\sigma _{r}\frac{\partial \Psi }{\partial \varphi })+i\frac{\omega _{so}}{%
\Omega }(\frac{\partial \overline{\Psi }}{\partial \varphi },\sigma _{r}\Psi
)+i\frac{\omega _{so}}{\Omega }(\overline{\Psi },\frac{\partial \sigma _{r}}{%
\partial \varphi }\Psi )\}.
\end{equation*}%
The resulting continuity equation takes the form 
\begin{equation*}
\frac{\partial \rho }{\partial t}=i\{\frac{\partial (\overline{\Psi },\frac{%
\partial \Psi }{\partial \varphi })-(\frac{\partial \overline{\Psi }}{%
\partial \varphi },\Psi )}{\partial \varphi }+i\frac{\omega _{so}}{\Omega }%
\frac{\partial (\overline{\Psi },\sigma _{r}\Psi )}{\partial \varphi }\}
\end{equation*}%
and the current $J$ is given by%
\begin{equation*}
J=\{(\overline{-i\frac{\partial \Psi }{\partial \varphi }+\frac{\omega _{so}%
}{2\Omega }\sigma _{r}\Psi },\Psi )+(\overline{\Psi },-i\frac{\partial \Psi 
}{\partial \varphi }+\frac{\omega _{so}}{2\Omega }\sigma _{r}\Psi )\}
\end{equation*}%
or%
\begin{equation*}
J=2\func{Re}\{(\overline{\Psi },-i\frac{\partial \Psi }{\partial \varphi }+%
\frac{\omega _{so}}{2\Omega }\sigma _{r}\Psi )\}=2\func{Re}\{\Psi ^{\dagger
}(-i\frac{\partial \Psi }{\partial \varphi }+\frac{\omega _{so}}{2\Omega }%
\sigma _{r}\Psi )\}.
\end{equation*}

b) Because the orientation of the coordinate system in the upper arm is
opposite to that in the lower arm, the current in the latter is given by $%
J_{low}^{\mu }(\varphi ^{\prime })=-J_{up}^{\mu }(\varphi =-\varphi ^{\prime
})$. The resulting forms of the two currents are given, respectively, by
Eqs. (16a) and (16b).

\begin{center}
{\Large Figure captions}
\end{center}

Fig. 1: Device geometry and the local coordinates ($x$, $x^{\prime }$, $%
\varphi $ and $\varphi ^{\prime }$) pertaining to different parts of the
ring.

\bigskip

Fig. 2: Dependence of the conductance $G$ on the Aharonov-Casher phase $%
\Delta_{AC}$ for different incident wave numbers $ka$ at zero temperature. $%
G $ is a periodic and even function of $ka$, hence $ka$ was considered only
in the interval $[0,1/2]$.

\bigskip

Fig. 3: Dependence of the zero-temperature coefficent $g_{0}$ on the
Aharonov-Casher phase $\Delta_{AC}$ for different wave numbers $ka$.

\bigskip

Fig. 4: Dependence of the coefficent $g_{T}$ on the Aharonov-Casher phase $%
\Delta_{AC}$ for different temperatures $T$ and different values of the
Fermi wave number $k_{F}a.$

\bigskip

Fig. 5: Dependence of the conductance $G$ on the Aharonov-Casher phase $%
\Delta_{AC}$ at different temperatures $T$ and Fermi wave numbers $k_{F}a.$

\bigskip

Fig. 6: Dependence of $G$ on the SOI strength $\alpha$ and $ka$ at zero
temperature; $\alpha_{0}$ is the value $10^{-11}\mathrm{eVm}$\textrm{.}

\bigskip

Fig. 7: Magnetocunductance for various values of $\alpha$, in units of $%
\alpha_{0}=10^{-11}\mathrm{eVm}$, and at low (a) and high (b) temperatures.

\bigskip

Fig. 8: Positions of extrema in the magnetoconductance oscillation as a
function of $\alpha $ at low (a) and high (b) temperature.

\bigskip

Fig. 9: Minimum-maximum conversions in the oscillations of the
magnetoconductance due to changes in the SOI strength $\alpha$ for two
values of the temperature.

\end{document}